\documentclass[prb,twocolumn,showpacs,preprintnumbers,amsmath,amssymb,supers criptaddress, prb, floatfix]{revtex4}
\usepackage{graphicx}% Include figure files
\usepackage{dcolumn}% Align table columns on decimal point
\usepackage{bm}% bold math

\begin{document}

\preprint{APS/123-QED}

\title{Revised crystal structure model of Li$_2$NH by neutron powder diffraction}% Force line breaks with \\

\author{Kenji Ohoyama}
 \email{ohoyama@imr.tohoku.ac.jp}
 \author{Yuko Nakamori}
\author{Shin-ichi Orimo}%
\author{Kazuyoshi Yamada}
\affiliation{%
Institute for Materials Research, Tohoku University, Sendai 980-8577, Japan
}%

\date{\today}% It is always \today, today,
             %  but any date may be explicitly specified

\begin{abstract}
We performed neutron powder diffraction experiments on lithium imide Li$_2$NH, and have proposed a revised crystal structure model.
Li$_2$NH has a face-centered cubic structure with a partially occupied hydrogen site.
Of the possible crystal structure models that represent the obtained data, the model with $F\bar{4}3m$ symmetry having hydrogen atoms at the $16e$ site, in which only one hydrogen atom randomly occupies one of the four hydrogen positions around a nitrogen atom, is most probable.
For this model, the distance between the nearest nitrogen and hydrogen atoms is 0.82(6)\,\AA, and the angle between H-N-H is 109.5$^{\circ}$, which are close to those of the lithium amide LiNH$_2$, indicating that the structural circumstances around nitrogen and hydrogen are similar in Li$_2$NH and LiNH$_2$.
\end{abstract}

% PACS, the Physics and Astronomy
                             % Classification Scheme.
%\keywords{Suggested keywords}%Use showkeys class option if keyword
                              %display desired
\maketitle

\section{\label{sec:introduction}Introduction}
In this paper, we report the results of crystal structure refinement of the lithium imide material Li$_2$NH using the neutron powder diffraction technique without deuteration, and propose a revised crystal structure model.

Recently, complex hydrides have become important as novel hydrogen storage materials, because they have a higher gravimetric hydrogen density in comparison with conventional hydrogen storage materials.~\cite{complex1,complex2,complex3,conv1}
Li-N-H materials are particularly interesting because these materials consist of light elements; thus, Li-N-H compounds hold promise as novel hydrogen storage materials with higher gravimetric hydrogen density.~\cite{Chen, Orimo1, Nakamori1, Orimo2}

The knowledge of the fundamental properties of practical hydrogen storage materials, especially the accurate positions of hydrogen and its crystallographical circumstances, that is, the accurate crystal structures, is indispensable from the point of view of controlling and improving their performance.

The target material of this paper Li$_2$NH, is thought to have a simple face-centered cubic ($fcc$) structure with $Fm\bar{3}m$ symmetry (Fig.1); the lithium atoms are located at the $8c$ site, the nitrogen atoms at the $4a$ site, and the hydrogen atoms at the $4b$ site of the $Fm\bar{3}m$ space group ($Z$ = 4).~\cite{Li2NH}
However, its structure appears to be peculiar because its characteristics differ from lithium amide LiNH$_2$, from which Li$_2$NH is prepared, in the following aspects: (i) the distance between the nearest nitrogen and hydrogen atoms, $d_{\rm N-H}$, of Li$_2$NH  ($\sim$2.5\,\AA) is much larger than $d_{\rm N-H}\sim$0.7 \AA~for LiNH$_2$~\cite{LiNH2} and $d_{\rm N-H}\sim$1.022 \AA ~for monomeric unsolvated LiNH$_2$,~\cite{LiNH2_2} (ii) the local structure around the nitrogen atom for Li$_2$NH is quite different from that for LiNH$_2$.~\cite{LiNH2,LiNH2_2}

Moreover, based on recent first-principles calculations for LiNH$_2$,~\cite{Orimo2} it is likely that the crystal structure of LiNH$_2$ is stabilized by the ionic bonding between the Li$^{+}$ cation and  [NH$_2$]$^{-}$ anion.
This suggests that, even for Li$_2$NH, the ionic bonding between the Li$^{+}$ cation and [NH]$^{2-}$ anion must be important.
Therefore, $d_{\rm N-H}$ for Li$_2$NH is probably sufficiently short, such that the [NH]$^{2-}$ anion is stabilized.

Thus, we expected the crystal structure model of Li$_2$NH in Fig.1 to be incorrect, because the positions of the hydrogen atoms in Fig.1 are debatable because the structure was determined by x-ray diffraction technique, which has a disadvantage in determining the accurate positions of hydrogen; Li$_2$NH must have a crystal structure with a shorter $d_{\rm N-H}$, and structural circumstances around nitrogen that are similar to those for LiNH$_2$.
\begin{figure}
\includegraphics[width=6cm]{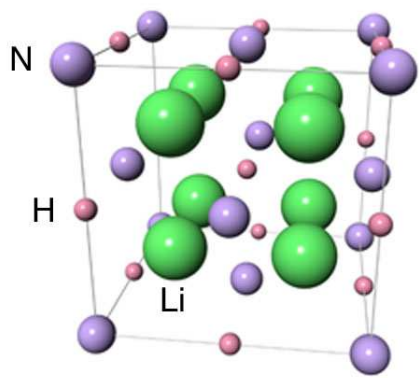}% Here is how to import EPS art
\caption{\label{fig:model}Crystal structure of Li$_2$NH which was reported in Powder Diffraction File.~\cite{Li2NH}.  The space group is $Fm\bar{3}m$.  One unit cell includes four molecules of Li$_2$NH.}
\end{figure}

To prove our expectation, we performed neutron diffraction experiments on a powder sample of Li$_2$NH, because the neutron diffraction technique can determine the positions of the hydrogen atoms in compounds more accurately, when compared with the x-ray diffraction technique.

\section{\label{sec:experiment}Experimental Details}
The powder sample of Li$_2$NH was prepared from the starting material LiNH$_2$ with purity exceeding 95\%, purchased from Aldrich Co. Ltd. by heat treatment at 623\,K for 24\,h in vacuum with a rotary pump.
To obtain direct information on the position of hydrogen itself, we did not deuterate the sample.
All the processes except the heat treatment, were done in a glove box with purified Ar (dew point below 183 K).
The purchased LiNH$_2$ included Li$_2$O as an impurity; we confirmed that the volume fraction of Li$_2$O was not changed by the heat treatment.

We performed the neutron powder diffraction experiments using the Kinken powder diffractometer for high efficiency and high resolution measurements, HERMES, belonging to the Institute for Materials Research, Tohoku University, installed at the reactor, JRR-3M, in the Japan Atomic Energy Research Institute, Tokai, Japan.~\cite{HERMES}
Neutrons with a wavelength of 1.8204\,\AA~ were obtained by the 331 reflection of the Ge monochromator and 12$'$-blank-sample-15$'$ collimation.
The powder sample of Li$_2$NH with a mass of 167\,mg was set in a vanadium cylinder with a diameter of 5\,mm, and sealed in a standard aluminium cell with purified Ar gas to avoid moisture.
For measurements at low temperatures, the sample was mounted on the cold head of a closed cycle He-gas refrigerator.
Since the thermal contact of Ar gas at low temperatures is not sufficient when compared with He gas,  we maintained the temperature of the sample at 10\,K for 48 h prior to the low temperature measurements.

For crystal structure refinement, we analyzed the obtained powder patterns using the Rietveld analysis program, RIETAN-2000.~\cite{RIETAN}

\begin{figure*}
\includegraphics[width=15cm]{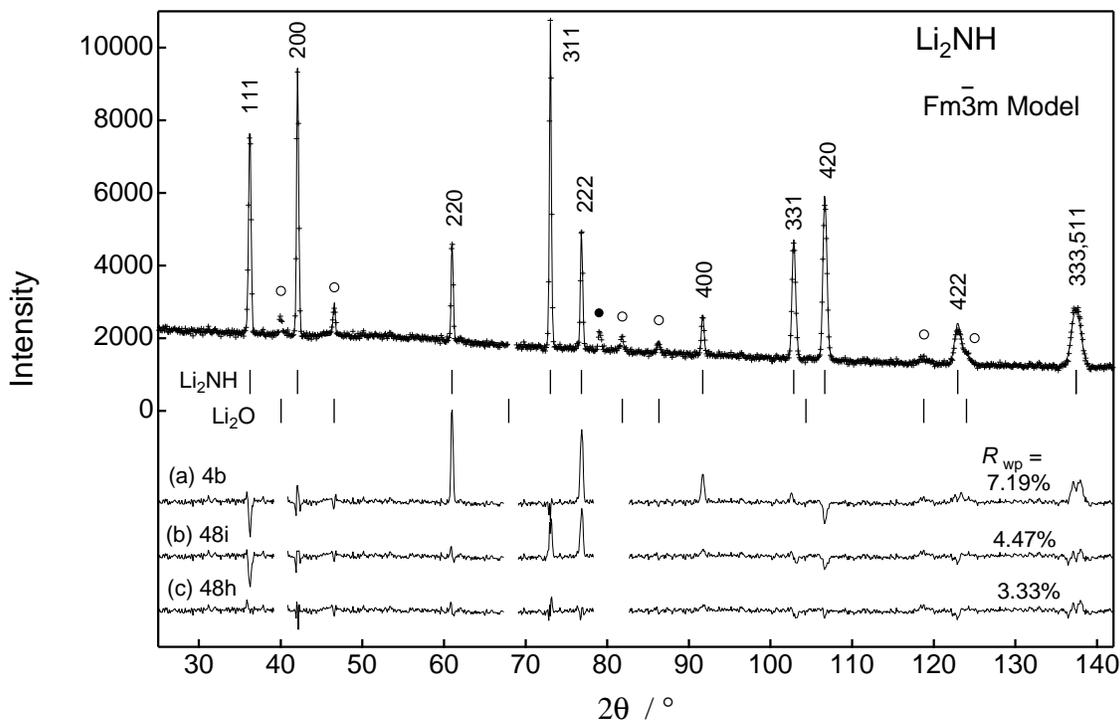}% Here is how to import EPS art
\caption{\label{fig:hermes_Fm3m}The powder pattern of Li$_2$NH at room temperature obtained on HERMES~\cite{HERMES} and some results obtained by Rietveld analysis for crystal structure models with $Fm\bar{3}m$ symmetry.  The solid lines represent the result of Rietveld fitting procedure by RIETAN-2000.~\cite{RIETAN}  The lines in the lower portion indicate the difference between the obtained data and calculation for the models in which the hydrogen atoms occupy (a) the $4b$ (Fig.1), (b) $48i$, and (c) $48h$ sites of the $Fm\bar{3}m$ space group. The conventional reliable factors, $R_{\rm wp}$, are also shown.  The details of the models are shown in the text.  The open circles indicate the Bragg peaks from Li$_2$O, while the peak at about 80$^{\circ}$ marked with the closed circle is unknown.}
\end{figure*}
\section{\label{sec:result}Results and Discussion}
Figure\,{\ref{fig:hermes_Fm3m}} shows the powder pattern of Li$_2$NH at room temperature obtained on HERMES and some results of Rietveld analysis for crystal structure models with $Fm\bar{3}m$ symmetry; the details of the models will be explained later.
The background originated due to incoherent scattering of hydrogen.
The small open circles indicate Bragg peaks for Li$_2$O that was included even in the starting material LiNH$_2$, while the peak marked with the closed circle is unknown; the peak position cannot be explained by those of possible impurities such as Li$_2$O, LiNH$_2$, Li$_3$N, LiOH, LiH, or Li.

First, we confirmed whether the crystal structure model in Fig.1 represents the data in Fig.2 or not.
Note that all the Bragg peaks of Li$_2$NH can be indexed with all even or all odd $h$, $k$, $l$ with only one lattice parameter, $a\sim$5.07\,\AA~, definitely indicating that Li$_2$NH has a $fcc$ structure as was believed so far.
However, we confirmed that the crystal structure model in Fig.1 cannot explain the pattern in Fig.2; the solid line (a) in the lower portion of Fig.2 is the difference between the obtained data and calculations for the model in Fig.1 in which the hydrogen atoms occupy the $4b$ site of the $Fm\bar{3}m$ space group.
Thus, we have concluded that the crystal structure model in Fig.1 is not correct, and must be revised as we expected.

To find the correct structure model of Li$_2$NH, we based the crystal structure refinement on the following two facts: (i) the structure has a $fcc$ symmetry with $a\sim$ 5.07\,\AA, (ii) one unit cell includes four molecules of Li$_2$NH.
In reality, it is easy to determine that there exist only five space groups that satisfy the two conditions: $F23$, $Fm\bar{3}$, $F432$, $F\bar{4}3m$, and $Fm\bar{3}m$.
However, we confirmed that no structure model with these symmetries represented the data in Fig.2, provided that the atoms fully or almost fully occupied each site.

Since no crystal structure model with fully occupied sites can represent the data, we propose a partially occupied model: the hydrogen atoms are located at a site with lower symmetry, for instance at the 192$l$ site of the $Fm\bar{3}m$ space group.
Since the multiplicity, $m$, for sites with lower symmetry is larger than four, the occupancy of the hydrogen site in this partially occupied model must be close to 4/$m$ to satisfy condition (ii).
In this model, we assumed that the nitrogen and lithium atoms exist at the same positions as shown in Fig.1 with 100\% occupancy, because the x-ray diffraction experiments, so far, did not disagree. Even for the partially occupied model, only the five space groups, $F23$, $Fm\bar{3}$, $F432$, $F\bar{4}3m$, and $Fm\bar{3}m$, can satisfy the conditions (i) and (ii).

For crystal structure refinement by the least-square fitting technique, some criterion for determining the  best model is required because the partially occupied model allows for greater freedom in fitting parameters, which yields a higher agreement automatically without any particular physical meaning.
Thus, to choose reasonable models, we judged the results of the fitting procedure according to the following criteria: when some models give nearly the same agreement, the model with the highest symmetry and/or the largest occupancy, in other words the smallest vacancy, is chosen as the best model.

We performed least-square fitting using the partially occupied models for $Fm\bar{3}m$.
The difference lines (b) and (c) in the lower portion of Fig.2 indicate the results for the models with the hydrogen atoms at the $48i$ and $48h$ sites, respectively.
The agreement of the model with the hydrogen atoms at the $48h$ site is good, while the model with the hydrogen atoms at $48i$ cannot represent the obtained pattern as indicated by the difference line (b) in Fig.2.
The solid line in the upper portion of Fig.2 is the calculation for the model with the hydrogen atoms at the $48h$ site.
The models with hydrogen atoms at lower symmetry sites, $96i$, $96k$, and $192l$, agree well as expected, while those with hydrogen atoms at higher symmetry sites than $48h$ and $48i$ cannot represent the data.
Thus, we can conclude that the best model for the $Fm\bar{3}m$ space group is the one with hydrogen atoms at the $48h$ site, in which the site occupancy of hydrogen is $\sim$4/48 = 8.3\%.

\begin{figure*}
\includegraphics[width=15cm]{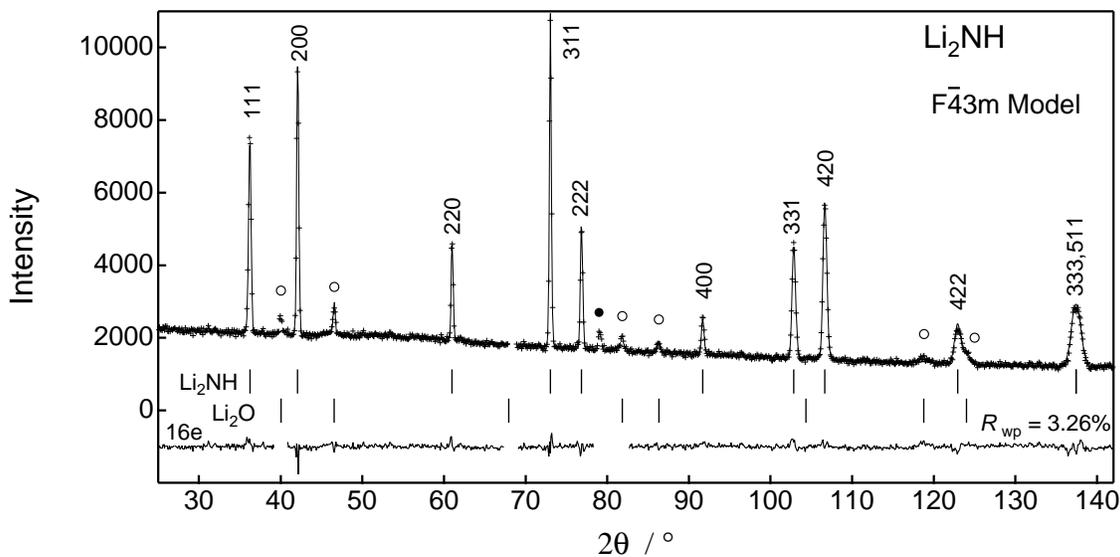}% Here is how to import EPS art
\caption{\label{fig:hermes_F43m}The best result of Rietveld analysis for the crystal model with $F\bar{4}3m$ symmetry having hydrogen atoms at $16e$.  The conventional reliable factor, $R_{\rm wp}$, is also shown.  The data is the same as that in Fig.2}
\end{figure*}

The partially occupied models with $F\bar{4}3m$ symmetry can also explain the data.
Figure 3 shows the best result of Rietveld analysis for the model in which the hydrogen atoms are located at the $16e$ site of the $F\bar{4}3m$ space group.
The models with the hydrogen atoms at the lower symmetry sites, $24f$, $24g$, $48h$, and $96i$, agree well as expected.
On the other hand, the model with hydrogen atoms at the  $4b$ site, which is the only site having  higher symmetry, cannot represent the data; in reality, this model is the same as the model for $Fm\bar{3}m$ with the hydrogen atoms at the $4b$ site, which cannot represent the data as shown in Fig.2.
Thus, we can conclude that the best model for $F\bar{4}3m$ is the one with the hydrogen atoms at the $16e$ site,  in which the site occupancy of hydrogen is $\sim$4/16 = 25\%.
 
Figure 4 shows the crystal structure models with hydrogen atoms at the $48h$ site under $Fm\bar{3}m$ symmetry (Model I), and at the $16e$ site under $F\bar{4}3m$ symmetry (Model II). The parameters obtained by Rietveld analysis are summarized in Table I.
As shown in Fig.4, a nitrogen atom is surrounded by twelve hydrogen positions for Model I, and by four hydrogen positions for Model II.

The other models with $F23$, $Fm\bar{3}$, and $F432$ symmetry can also represent the data; however, the best models for the three space groups essentially yield the same structures as those in Fig.4.
Thus, either Model I or II in Fig.4 can be considered to be the correct model, though the results of the present neutron diffraction experiments cannot distinguish the two models in Fig.4, because these exhibit nearly the same agreement between the data.
\begin{figure}
\includegraphics[width=7.5cm]{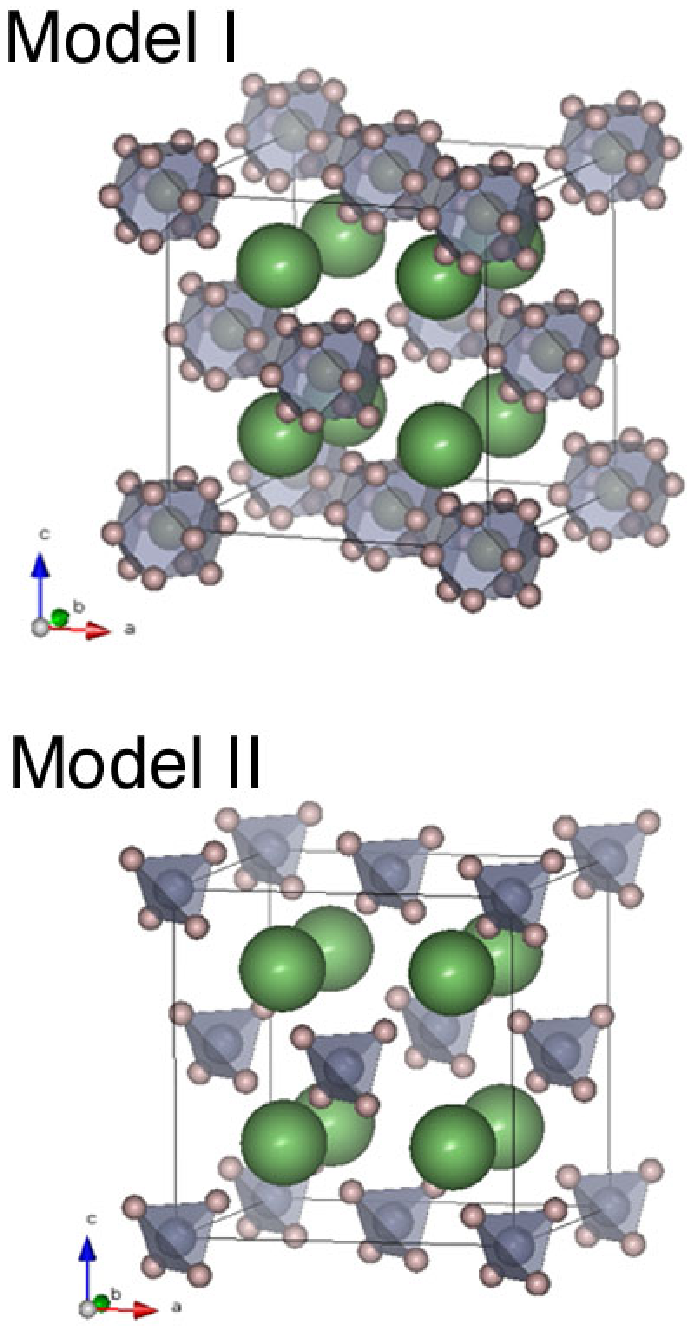}% Here is how to import EPS art
\caption{\label{fig4}Crystal structure models of Li$_2$NH. Model I: $Fm\bar{3}m$ symmetry for the hydrogen atoms at the $48h$ site (0, $y$, $y$), determined from the analysis in Fig.2, Model II: $F\bar{4}3m$ symmetry for hydrogen atoms at the $16e$ site ($x$, $x$, $x$), determined from the analysis in Fig.3.  The parameters are summarized in Table I.  These figures were drawn with VENUS developed by Dilanian and Izumi.}
\end{figure}
\begin{table}
\caption{\label{tab:table1}Summary of crystal structure parameters of Model I and Model II.  the site occupancy is denoted by $g$.   $d_{\rm H-N}$, $d_{\rm N-Li}$, $d_{\rm H-H}$, and $\theta_{\rm H-N-H}$ are the distance between the nearest nitrogen and hydrogen atoms, between the nearest nitrogen and lithium atoms, between the nearest hydrogen atoms, and the angle of H-N-H, respectively.  $R_{\rm wp}$ and $R_{\rm e}$ are the conventional reliable factors.~\cite{RIETAN}}
\begin{ruledtabular}
\begin{tabular}{cccccccc}
$a$ (\AA)&5.0769(1)&&&&\\
$V$ (\AA$^3$)&130.860(5)&&&&\\
\hline
\\
\\
Model I \\
Space Group&$Fm\bar{3}m$ &&&&\\
\\
&Li&N&H\\
\hline
site&$4c$&$4a$&$48h$\\
$g$&1.0&1.0&0.087(5)\\
$x$&0.25&0&0\\
$y$&0.25&0&0.110(3)\\
$z$&0.25&0&0.110(3)\\
$B$ (\AA$^2$)&6.5(5)&2.3(3)&1.86(1)\\
\\
\hline
$d_{\rm H-N}$ (\AA)&&0.79(2)&&&\\
$\theta_{\rm H-N-H}$ (deg.)&&60$^{\circ}$\footnotemark[1]&&&\\
$d_{\rm H-H}$ (\AA)&&0.79(2)&&&\\
$d_{\rm N-Li}$ (\AA)&&2.19836(5)&&&\\
$R_{\rm wp}$&&3.33\%&&&\\
$R_{\rm e}$&&2.30\%&&&\\
\hline
\\
\\
\\
Model II \\
Space Group&$F\bar{4}3m$&&&&\\
\\
&Li1&Li2&N&H\\
\hline
site&$4c$&$4d$&$4a$&$16e$\\
$g$&1.0&1.0&1.0&0.27(1)\\
$x$&0.25&0.75&0&0.093(6)\\
$y$&0.25&0.75&0&0.093(6)\\
$z$&0.25&0.75&0&0.093(6)\\
$B$ (\AA$^2$)&5.5(5)\footnotemark[2]&5.5(5)\footnotemark[2]&1.8(3)&7(1)\\
\\
\hline
$d_{\rm H-N}$ (\AA)&0.82(6)&&&\\
$\theta_{\rm H-N-H}$ (deg.)&109.47$^{\circ}$\footnotemark[1]&&&\\
$d_{\rm H-H}$ (\AA)&1.34(4)&&&\\
$d_{\rm N-Li}$ (\AA)&2.19836(5)&&&\\
$R_{\rm wp}$&3.26\%&&&\\
$R_{\rm e}$&2.30\%&&&\\
\end{tabular}
\footnotetext[1]{$\theta_{\rm H-N-H}$ is exactly the same as the angle between the [011] and [101] directions for Model I,  and the angle between the [111] and [$\bar{1}\bar{1}1$] directions for Model II, for instance.}
\footnotetext[2]{$B$ of Li1 and Li2 were assumed to be equal.}
\end{ruledtabular}
\end{table}

By comparing the two models in Fig.4, we have concluded that the Model II, in which the hydrogen atoms are located at the $16e$ site of the $F\bar{4}3m$ symmetry is the better model, because of the following reasons: (i) the site symmetry of the $16e$ site of $F\bar{4}3m$ symmetry is higher than that of $48h$ of $Fm\bar{3}m$ symmetry, (ii) therefore, Model II yields three times higher occupancy, in other words, a smaller vacancy, of the hydrogen site than in Model I.
In fact, Model II yields the slightly better reliable factor, $R_{\rm wp}$, than Model I.

For Model II, a nitrogen atom is surrounded by four hydrogen positions.
However, it is also notable that the occupancy of the hydrogen site is approximately 1/4.
Therefore, in the local structure of Li$_2$NH, only one among the four hydrogen positions around a nitrogen atom is randomly occupied by a real hydrogen atom; consequently, the local structure is as shown in Fig.5.
Spatial and/or time average of the local structure in Fig.5 yield the average structure in Fig.4 observed by neutron diffraction.

\begin{figure}
\includegraphics[width=6cm]{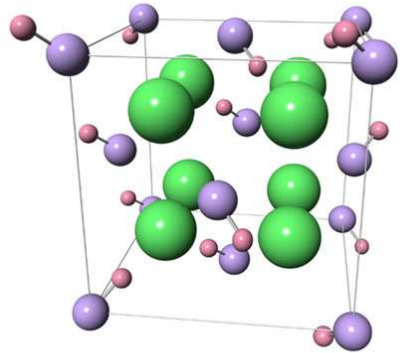}% Here is how to import EPS art
\caption{\label{fig5}Local structure of a unit cell of Li$_2$NH based on Model II in Fig.4.  Only one of the four hydrogen positions around a nitrogen atom in Model II is randomly occupied.}
\end{figure}

Let us compare the spatial relation between the nearest nitrogen and hydrogen atoms of Li$_2$NH and LiNH$_2$.
In the average structure of Li$_2$NH for Model II shown in Fig.4, $d_{\rm N-H}$ is 0.82(6)\,\AA, and the angle between H-N-H is 109.5$^{\circ}$, which are close to $d_{\rm N-H}$ = 1.022\,\AA~ and $\theta_{\rm H-N-H}$ = 106.9$^\circ$ for monomeric unsolvated LiNH$_2$,~\cite{LiNH2_2} and $d_{\rm N-H}$ = 1.02\,\AA~ and $\theta_{\rm H-N-H}$ = 106.6$^\circ$ for ammonia, in comparison with those in the model in Fig.1.
Thus, this result indicates that the spatial relation of nitrogen and hydrogen in the average structure of Li$_2$NH is basically the same as that of LiNH$_2$, as we expected.
However, it should be noted that $d_{\rm N-H}$ for Li$_2$NH is 20\% shorter than those for LiNH$_2$ and ammonia.
Since $d_{\rm N-H}$ for LiNH$_2$ and ammonia are approximately consistent with the sum of the covalent bond radius of nitrogen and hydrogen,  the shorter $d_{\rm N-H}$ for Li$_2$NH implies that the N-H bonding in Li$_2$NH cannot be understood by the normal covalent bonding.

This poses a problem if the structure is stable, because the local structure in Fig.5 includes a huge vacancy of 75\% at the hydrogen site. 
If not, then a structural phase transition probably occurs at lower temperatures.
However, the results of measurements at lower temperatures showed no structural phase transition at least above approximately 10\,K.

%%%%
The fact that the characteristic structure in Fig.5 exists at room temperature indicates that it is probable that a hydrogen atom hops between hydrogen positions around a nitrogen atom to satisfy both the crystallographical symmetry of the average structure and electroneutrality.
Since such a hopping yields the characteristic dynamic properties, inelastic neutron scattering must be important to understand the stability of hydrogen in Li$_2$NH.

In conclusion, the crystal structure model of the lithium imide Li$_2$NH has been revised; Li$_2$NH has a $fcc$ structure with a partially occupied hydrogen site.
Of the two possible structure models with $Fm\bar{3}m$ symmetry having hydrogen atoms at the $48h$ site (Model I in Fig.4), and $F\bar{4}3m$ with the hydrogen atoms at the $16e$ site (Model II in Fig.4), the latter model is more probable because it is relatively simple.
In the model, only one hydrogen atom randomly occupies one of the four hydrogen positions around a nitrogen atom.
In the average structure, the crystallographical circumstances around hydrogen and nitrogen are similar to that of LiNH$_2$, from which Li$_2$NH is prepared.

To distinguish the two possible models experimentally, high resolution neutron powder diffraction experiments with a wide momentum transfer range are needed.
Moreover, nuclear magnetic resonance and neutron inelastic scattering studies of Li$_2$NH  are in progress to understand the dynamics of hydrogen.

\begin{acknowledgments}
The authors would like to thank Mr. K. Nemoto of IMR, Tohoku University for his helpful assistance in the neutron scattering experiments.
This work was partially supported by the Ministry of Education, Culture, Sports, Science and Technology, ``Grant-in-Aid for Encouragement of Young Scientists (A), \#15686027," and by the New Energy and Industrial Technology Development Organization (NEDO), ``Basic Technology Development Project for Hydrogen Safety and Utilization (2003-2004), \#03001387."
\end{acknowledgments}

\newpage %Just because of unusual number of tables stacked at end
%\bibliography{LiNH}% Produces the bibliography via BibTeX.

\end{document}